\documentclass[10pt]{article}

\usepackage{amsmath}
\usepackage{amsthm}
\usepackage{amssymb}
\usepackage{amsfonts}
\usepackage{revsymb}
\usepackage{graphicx}
\usepackage{longtable}

\newtheorem{lemma}{Lemma}[section]
\newtheorem{theorem}[lemma]{Theorem}

\theoremstyle{definition}
\newtheorem{definition}[lemma]{Definition}
\theoremstyle{remark}

\newcommand{\ket}[1]{{|#1\rangle}}
\newcommand{\bra}[1]{{\langle#1|}}

\newcommand{\du}[1]{{#1^\dag}}

\newcommand{\abs}[1]{|#1|}

\newcommand{\oper}[1]{#1}
\newcommand{\op}[1]{#1}

\newcommand{\sop}[1]{\mathbf{\hat{#1}}}
\newcommand{\id}{{\oper{\openone}}}

\newcommand{\set}[1]{{\mathbf #1}}

\def\H{{\mathcal H}}

\newcommand\be{\begin{eqnarray}}
\newcommand\ee{\end{eqnarray}}
\newcommand\ba{\begin{array}}
\newcommand\ea{\end{array}}

\begin{document}
\title{Limits and restrictions of private quantum channel}
\author{Jan Bouda${}^{1}$ and Mario Ziman${}^{1,2}$\\
\\
\small  ${}^{1}$
Faculty of Informatics, Masaryk University,\\
\small Botanick\'a 68a, 602 00 Brno, Czech Republic\\\\
\small ${}^{2}$
Research Center for Quantum Information, Slovak Academy of Sciences\\
\small Dubravsk\'a cesta 9, 845 11 Bratislava, Slovakia
}

%\date{\today}
\maketitle

\begin{abstract}
We study private quantum channels on a single qubit, which encrypt given set of plaintext states $\set{P}$. Specifically, we determine all achievable states $\op{\rho}^{(0)}$ (average output of encryption) and for each particular set $\set{P}$ we determine the entropy of the key necessary and sufficient to encrypt this set. It turns out that single bit of key is sufficient when the set $\set{P}$ is two dimensional. However, the necessary and sufficient entropy of the key in case of three dimensional $\set{P}$ varies continuously between $1$ and $2$ bits depending on the state $\op{\rho}^{(0)}$. Finally, we derive private quantum channels achieving these bounds. We show that the impossibility of universal NOT operation on qubit can be derived from the fact that one bit of key is not sufficient to encrypt qubit.
\end{abstract}

\section{Introduction}
Quantum cryptography \cite{Gisin+Ribordy...-Quant_Crypt:2001,Bouda-Encryptionofquantum-2004} (for a popular review see \cite{Gottesman+Lo-From_quant_cheat:2000}) is a rapidly developing branch of quantum information processing. The results of quantum cryptography include quantum key distribution \cite{Bennett+Brassard-Quant_crypt:1984,Ekert-Quant_crypt_based:1991}, quantum secret sharing \cite{Hillery+Buzek...-Quant_secre_sharing:1999,Cleve+Gottesman...-share_quant_secre:1999}, quantum oblivious transfer \cite{Bennett+Brassard...-Pract_quant_obliv:1991,Crepeau-Quant_obliv_trans:1994} and other cryptographic protocols \cite{Gruska-Quant_compu:1999}. Quantum cryptography has two main goals: solutions to classical cryptographic primitives, and quantum cryptographic primitives.

The first goal is to design solutions of cryptographic primitives, which achieve a higher (provable) degree of security than their classical counterparts. The degree of security should be better than the security of any known classical solution, or it should be of the degree that is even not achievable by using classical information theory at all. Another alternative is to design a solution which is more efficient\footnote{According to time, space or communication complexity.} than any classical solution of comparable security.

The second class of cryptosystems is motivated by the evolution of applications of quantum information processing, regardless whether their purpose is cryptographic, communication complexity based or algorithmic. These cryptosystems are designed to manipulate quantum information. As applications of quantum information processing start to challenge a number of their classical counterparts, the need to secure quantum communications in general is getting more urgent. Therefore, there is a large class of quantum primitives which should secure quantum communication in the same way as classical communication is secured. These primitives include encryption of quantum information using both classical \cite{Ambainis+Mosca...-Priva_quant_chann:2000,Boykin+Roychowdhury-Optim_encry_quant:2000,Oppenheim.Horodecki-Howtoreuse-2003} and quantum key \cite{Leung-Quant_Verna_ciphe:2000}, authentication of quantum information \cite{Barnum+Crepeau...-Authenticatio_of_q_mes:2002}, secret sharing of quantum information \cite{Cleve+Gottesman...-share_quant_secre:1999,Gottesman-theor_quant_secret:1999}, quantum data hiding \cite{DiVincenzo+Hayden...-Hidin_quant_data:2002} and even commitment to a quantum bit \cite{Bouda-Encryptionofquantum-2004}, oblivious transfer of quantum information \cite{Bouda-Encryptionofquantum-2004} and others.

In this paper we concentrate on the encryption of quantum information with classical key \cite{Ambainis+Mosca...-Priva_quant_chann:2000} described and explained in Section \ref{sec:Private-quantum-channel}. At the end of Section \ref{sec:Private-quantum-channel} we introduce notation and one theorem we will be using through the remaining sections. To begin our analysis, in Section \ref{sec:Achievable-states} we investigate for a given set $\set{P}$ the set of all possible states $\op{\rho}^{(0)}$ such that there exists a private quantum channel $\sop{E}$ with the property $\forall\op{\rho}\in\set{P}:\sop{E}(\op{\rho})=\op{\rho}^{(0)}$ and we determine that it forms a ball within the Bloch sphere centered in $\frac{1}{2}\id$. In the Sections \ref{sub:General-analysi}--\ref{sec:Realiza-entropy} we derive all possible private quantum channels for a given set $\set{P}$ and state $\op{\rho}^{(0)}$ and analyze necessary and sufficient entropy of the key of such PQC. We also explicitly construct PQCs achieving this bound. Another interesting result contained in Section \ref{sec:Realiza-entropy} is that any PQC encrypting given set $\set{P}$ of two linearly independent states encrypts also any two-dimensional set $\set{P}'$ parallel to $\set{P}$ and lying in the plane spanned by $\set{P}$ and $\frac{1}{2}\id$.

We conclude our paper in Section \ref{sec:Conclus} by few comments on possible generalizations of the described techniques to systems of higher dimension.

%%%%%%%%%%%%%%%%%%%%%%%%%%%%%%%%%%%%%%%%%%%%%%%%%%%%%%%%%%%%%%%%%%%%%%%%%%%%%%%%%%%%%%%%%%%%%%%%%%%%%%%%%%%%%%%%%%%%%%%%
\section{Private quantum channel}
\label{sec:Private-quantum-channel}
%%%%%%%%%%%%%%%%%%%%%%%%%%%%%%%%%%%%%%%%%%%%%%%%%%%%%%%%%%%%%%%%%%%%%%%%%%%%%%%%%%%%%%%%%%%%%%%%%%%%%%%%%%%%%%%%%%%%%%%%

The private quantum channel \cite{Ambainis+Mosca...-Priva_quant_chann:2000} is a general framework designed to
perfectly encrypt an arbitrary quantum system using a classical key.
\begin{definition}
\label{def:PQCdef} Let $\set{P}\subseteq{\mathcal S}({\H}_{1,\dots,n})$ be a set of $n$-qubit states\footnote{To make the definition
easier we work with qubits. To obtain an equivalent definition for arbitrary quantum systems $A$ it suffices to replace
$\H_{1,\dots,n}$ by $\H_A$ and $\H_{n+1,\dots,m}$ by $\H_{anc}$.}, $\sop{E}=\{(p_i,\op{U}_i)\}_i$ be a superoperator,
where each $\oper{U}_i$ is a unitary operator on $\H_{1,\dots,m},n\le m$, $p_i\geq1$ and $\sum_i p_i=1$. Let
$\oper{\rho}_{anc}$ be an $(m-n)$ qubit density matrix and $\oper{\rho}^{(0)}$ be an $m$-qubit density matrix. Then
$[\set{P},\sop{E},\oper{\rho}_{anc},\oper{\rho}^{(0)}]$ is a {\bf private quantum channel (PQC)}\index{private quantum
channel}\index{PQC} if and only if for all $\op{\rho}\in \set{P}$ it holds that
\begin{equation}
\label{equ:PQC_definition}
\sop{E}(\op{\rho}\otimes\oper{\rho}_{anc})=\sum_i p_i\oper{U}_i
(\op{\rho}\otimes\oper{\rho}_{anc})\du{\oper{U}_i}=\oper{\rho}^{(0)}.
\end{equation}
\end{definition}

The definition of the private quantum channel establishes the following cryptosystem. Alice wants to establish a communication (quantum) channel with Bob with the property that any state $\op{\rho}\in\set{P}$ will be transmitted securely. The security in this case means that Eve gets no advantage (information) by intercepting the transmitted message.

\begin{figure}
  \begin{center}
  \includegraphics[width=8cm]{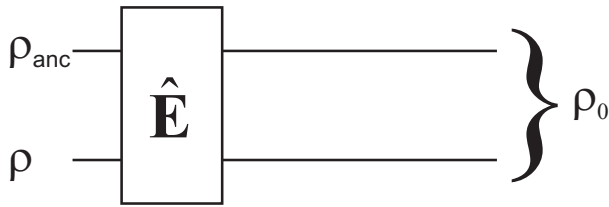}
  \caption{Encryption using PQC.}
  \label{fig2}
  \end{center}
\end{figure}

The encryption of the plaintext is done in the way that one operator, chosen randomly out of the operators $\{\oper{U}_i\}_i$, is applied to the plaintext system. The operator $\oper{U}_i$ is chosen with probability $p_i$. The classical key specifies which of the unitary operators was applied. The unitary operators $\oper{U}_i$ are acting on $\H_{1,\dots,m}$, while the state $\op{\rho}$ is only $n$-qubit state. The encryption operation $\oper{U}_i$ is performed on the Hilbert space $\H_{1,\dots,m}$, the plaintext space $\H_{1,\dots,n}$ is a subspace of $\H_{1,\dots,m}$. The encryption operation is defined on the (possibly) larger space than the plaintext to allow optional encryption of the plaintext together with an ancillary system. The encryption operators, therefore, act on a tensor product of the plaintext Hilbert space $\H_{1,\dots,n}$ and the ancillary Hilbert space $\H_{n+1,\dots,m}$, which is originally factorized (decoupled) from the plaintext. The ancillary Hilbert space is initially in the state $\oper{\rho}_{anc}$ (see Figure \ref{fig2}).

%It is usually possible to ignore the ancilla when deriving general results, since the PQC $[\set{P},\sop{E},\op{\rho}_{anc},\op{\rho}^{(0)}]$, $\set{P}=\{\op{\rho}_i\}_i$, is equivalent to the PQC
%$[\set{P}',\sop{E},\op{\rho}^{(0)}]$, where $\set{P}'=\{\op{\rho}_i\otimes\op{\rho}_{anc}|\op{\rho}_i\in\set{P}\}$.
%\index{private quantum channel}\index{PQC} It serves rather when designing a PQC for a concrete situation. We will also use the set $\set{P}$ as a set of all plaintext states, which should be encrypted. Namely, we will accept density operators as members of the set $\set{P}$.

The security of the scheme can be explained in the following way: Without knowledge of the key (i.e. without specific knowledge about which of the operators was used) any initial state $\op{\rho}\in \set{P}$ together with the ancilla appears to be in the state $\oper{\rho}^{(0)}$ after the encryption. The state $\oper{\rho}^{(0)}$ is the same for all $\op{\rho}\in \set{P}$, it is independent of the input state. It means that all states from the set $\set{P}$ are physically indistinguishable after the encryption.

A dual point of view of the security is also possible.
Let us denote by $\set{C}=\sop{E}[\set{P}]$ the set of all ciphertexts, i.e. $\op{\rho}^{(c)}_i=\op{U}_i\op{\rho} \du{\op{U}_i} \in\set{C}$
for each encryption operation $\op{U}_i$ and plaintext state $\op{\rho}\in\set{P}$. The encryption key (represented by the sequence $i_1,\dots,i_n$) is used also for the decryption. The only difference in the case of the decryption
is that the inverse operations $\du{\op{U}_i}$ are applied, i.e. $\op{\rho}_i^{(c)}\mapsto \du{\op{U}_i}\op{\rho}_i^{(c)} \op{U}_i=\op{\rho}\in\set{P}$. Formally the decryption procedure induces a transformation $\sop{D}[\op{\rho}]=\sum_i p_i \du{\op{U}_i} \op{\rho} \op{U}_i$. It describes the result of a decryption of a particular ciphertext without knowledge which key was used to encrypt it. The probabilities $\{p_i\}_i$ are the same as in the case of Eq.
\eqref{equ:PQC_definition}, because the probability that the key $\op{U}_i$ was used is $p_i$.
In general, each encryption operation $\op{U}_i$ defines a different set of ciphertexts $\set{C}_i$. The linear span $\overline{\set{C}}_i$ is just rotated set of plaintexts $\overline{\set{P}}$.

From Eq. \eqref{equ:PQC_definition} it follows that the information about the ciphertext contained in the plaintext is $I(\oper{\rho}_P:\oper{\rho}_C)=0$. However, from the symmetry of the mutual information it follows that the information about plaintext contained in the ciphertext is also $0$, therefore, the dual equation also holds
\begin{equation}
\label{equ:PQC_definition_dual}
\sop{D}(\oper{\rho})=\sum_i p_i\du{\oper{U}}_i
(\oper{\rho}){\oper{U}_i}=\oper{\rho}^{(1)},
\end{equation}
where $\oper{\rho}^{(1)}$ is fixed for all ciphertext states $\op{\rho}$. The superoperator $\sop{D}=\{(p_i,\du{\op{U}}_i)\}_i$ is not an inverse of the superoperator $\sop{E}$ in the standard meaning. It describes the result of a decryption of a particular ciphertext without knowledge which key was used to decrypt it.

From the mathematical point of view the encryption transformation $\sop{E}$ is defined as a convex combination of unitary maps. Using the ancilliary system the encryption can be still defined only in terms of the system under consideration. Tracing out the ancilla we obtain a map $\sop{E}_s$ with the action defined by $\sop{E}_s[\op{\rho}]={\rm Tr}_{anc}\sop{E}[\op{\rho}\otimes\op{\rho}_{anc}]$. The usage of ancilla results in most general form of the quantum channel, i.e. $\sop{E}_s[\op{\rho}]=\sum_i p_i \sop{G}_i[\op{\rho}]$ with $\sop{G}_i[\op{\rho}]={\rm Tr}_{anc} [\op{U}_i\op{\rho}\otimes\op{\rho}_{anc} \du{\op{U}}_i]$. However, for the decryption the ancilliary system is necessary.
In this paper we will analyze PQC without additional ancillas, so the encryption is formally a convex combination of unitary transformations. It means the encryption is described by a unital completely positive map, i.e. it preserves the total mixture $\frac{1}{2}\id$.

Finally, we will introduce one definition and one theorem, which will be used through this paper.
\begin{definition}
Let $\set{P}=\{\op{\rho}_i|i\in\set{I}\}$, where $\set{I}$ is an index
set. We define the set $\overline{\set{P}}$ as
\begin{equation}
\overline{\set{P}}=\left\{\op{\rho}=\sum_{i\in\set{I}}\lambda_i\op{\rho}_i\Bigg|\op{\rho}_i\in\set{P},\lambda_i\in\mathbb R,\sum_{i\in\set{I}}\lambda_i=1\right\}.
\end{equation}
Especially in the case when $\set{P}=\{\op{\rho}_1,\op{\rho}_2\}$ the set $\overline{\set{P}}$ contains all operators of the form $\lambda\op{\rho}_1+(1-\lambda)\op{\rho}_2$.
\end{definition}
From now on we will denote the maximally mixed state\footnote{I.e. the state nearest to the maximally mixed state $\frac{1}{2}\id$ according to the trace distance. It is also the nearest state in the Bloch ball.} in $\overline{\set{P}}$ as $\overline{\rho}$.

\begin{theorem}
\label{the:PQC_linear_span}
Let $[\set{P},\sop{E},\op{\rho}^{(0)}]$ be a PQC. Then $\sop{E}(\op{\rho})=\op{\rho}^{(0)}$ for any operator $\op{\rho}\in\overline{\set{P}}$. Note that we are interested only in operators $\op{\rho}$ with nonnegative eigenvalues, since the operators with negative eigenvalue(s) are not valid quantum states.
\end{theorem}
\begin{proof}
The proof follows from the linearity of $\sop{E}$.
\end{proof}

%%%%%%%%%%%%%%%%%%%%%%%%%%%%%%%%%%%%%%%%%%%%%%%%%%%%%%%%%%%%%%%%%%%%%%%%%%%%%%%%%%%%%%%%%%%%%%%%%%%%%%%%%%%%%%%%%%%%%%
\section{Achievable states $\op{\rho}^{(0)}$}
\label{sec:Achievable-states}
%%%%%%%%%%%%%%%%%%%%%%%%%%%%%%%%%%%%%%%%%%%%%%%%%%%%%%%%%%%%%%%%%%%%%%%%%%%%%%%%%%%%%%%%%%%%%%%%%%%%%%%%%%%%%%%%%%%%%%

In this section we will derive several results on PQC on a single qubit. Our first question is 'What are the possible states $\op{\rho}^{(0)}$ given a specific set $\set{P}$?'. In \cite{Ambainis+Mosca...-Priva_quant_chann:2000} it was proved that
the only possible candidate for the state $\op{\rho}^{(0)}$ is $\frac{1}{2}\id$, whenever $\frac{1}{2}\id$ can be expressed as a convex combination of states from $\set{P}$. We will generalize this result for any set $\set{P}$ to calculate the minimal entropy of the key necessary and sufficient to encrypt a specific set $\set{P}$.

All information about the action of $\sop{E}$ we have is its behaviour on the set of plaintexts $\set{P}$. Each element of this set is transformed into the fixed state $\op{\rho}^{(0)}$. This determines the channel $\sop{E}$ completely,
or incompletely depending on the set $\set{P}$. However, in the case of incomplete specifications the choice of the channel $\sop{E}$ has no impact on the security. Under the action of the channel $\sop{E}$ each operator $\op{\rho}\in\overline{\set{P}}$ is transformed into $\op{\rho}^{(0)}$. This follows directly from the linearity of the transformation $\sop{E}$.

Let us assume a PQC given by operators $\{(p_i,\op{U}_i)\}_i$ with some $\op{\rho}^{(0)}$. By applying the unitary transformations $\op{U}_i^\prime = \op{V} \op{U}_i$ ($\op{V}$ is unitary) we obtain that $\sum_i p_i \op{U}_i^\prime \op{\rho} \op{U}_i^{\prime\dagger}= \sum_i p_i \op{V} \op{U}_i\op{\rho} \du{\op{U}_i} \du{\op{V}} = \op{V}\op{\rho}^{(0)} \du{\op{V}}=\op{\rho}^{\prime(0)}$ is fixed for all plaintext states. It follows that the unitarily transformed PQC is again a PQC  with unitarily transformed average output state. Moreover,
a convex combination of two PQC channels $\sop{E}_1,{\sop{E}}_2$ for the given set
$\set{P}$ is again a PQC channel for $\set{P}$. In particular, $\sop{E}=\pi_1\sop{E}_1+\pi_2\sop{E}_2$ is PQC with
$\op{\rho}^{(0)} = \pi_1\op{\rho}^{(0)}_1+\pi_2\op{\rho}^{(0)}_2$, i.e. $\sop{E}[\op{\rho}]=\op{\rho}^{(0)}$ for all $\op{\rho}\in\set{P}$. Thus, for a given set of plaintexts $\set{P}$ the set of all possible private quantum
channels is convex. The set of achievable states is convex as well. It is formed by orbits of states under the action of the whole unitary group.

For any TCP map $\sop{E}$ the following inequality holds
\begin{equation}
\label{equ:contractive_superoperator}
D(\op{\rho},\op{\sigma})\geq D(\sop{E}(\op{\rho}),\sop{E}(\op{\sigma}))
\end{equation}
for the  distance measure $D(\op{\rho},\op{\sigma})={\rm Tr}|\op{\rho}-\op{\sigma}|$ on mixed states, i.e. two quantum states $\op{\rho},\op{\sigma}$ cannot become more distinguishable after applying a TCP transformation. We have already mentioned that we consider that the encryption superoperator $\sop{E}$ is unital (we do not consider the ancilla here) and therefore from Eq. \eqref{equ:contractive_superoperator} we have
\begin{equation}
D\left(\op{\rho},\frac{1}{2}\id\right)\geq D\left(\op{\rho}^{(0)},\frac{1}{2}\id\right),
\end{equation}
where $\op{\rho}\in\overline{\set{P}}$ is any state in the set $\overline{\set{P}}$ and $\op{\rho}^{(0)}=\sop{E}(\op{\rho})$ is fixed for all states $\op{\rho}\in\overline{\set{P}}$. We use the fact that for unital maps $\sop{E}[\frac{1}{2}\id]=\frac{1}{2}\id$. Especially this equation holds for the state $\overline{\rho}$, which is the most mixed density operator in $\overline{\set{P}}$ (the nearest point to $\frac{1}{2}\id$ in the Bloch ball).

Therefore the condition is that given the set of plaintext states $\overline{\set{P}}$ any achievable state $\op{\rho}^{(0)}$ fulfills the condition
\begin{equation}
\label{equ:rho_0_condition}
D\left(\overline{\rho},\frac{1}{2}\id\right)\geq D\left(\op{\rho}^{(0)},\frac{1}{2}\id\right).
\end{equation}
As a consequence, we obtain the result of \cite{Ambainis+Mosca...-Priva_quant_chann:2000}
that the state $\op{\rho}^{(0)}=\frac{1}{2}\id$ whenever $\frac{1}{2}\id$ is contained in the convex span of the set $\set{P}$.

Provided that the most mixed state $\overline{\rho}$ in
$\overline{\set{P}}$ is not $\frac{1}{2}\id$, the state
$\op{\rho}^{(0)}$ must have the same or a smaller distance from
$\frac{1}{2}\id$ than $\overline{\rho}$. This is the necessary
condition each candidate to the state $\op{\rho}^{(0)}$ must obey. In
this sense the set of potential candidates $\op{\rho}^{(0)}$ forms a
ball within the Bloch ball, with the center in $\frac{1}{2}\id$ and
the radius given by the distance of $\overline{\rho}$ and
$\frac{1}{2}\id$. Let us denote this ball (set of allowed states) by
$b$. This condition is necessary, it remains to verify whether it is
also sufficient, i.e. whether for a given $\set{P}$ and $\forall\
\op{\rho}^{(0)}\in b$ there exists a suitable TCP superoperator
$\sop{E}_{\op{\rho}^{(0)}}$, or equivalently whether the set of all achievable states coincides with those allowed by  inequality \eqref{equ:rho_0_condition}.

In what follows we will analyze the achievability of $\op{\rho}^{(0)}$. Let us first consider two trivial cases. When $\set{P}$ has only a single member, then there is nothing to encrypt. If the set $\set{P}$ contains at least four linearly independent members, then $\overline{\set{P}}$ already spans the whole Bloch ball and $\frac{1}{2}\id\in\overline{\set{P}}$. It follows that the PQC maps all states to $\frac{1}{2}\id$, so the set of achievable states contains only single element. In subsequent sections we will analyze the remaining two cases:
i) set $\overline{\set{P}}$ is two-dimensional, and ii) set $\overline{\set{P}}$ is three-dimensional.

In Sections
\ref{sub:General-analysi}--\ref{sub:Three-linearl-indepen-states} we
will adopt analytical approach to prove that for all states
$\op{\rho}^{(0)}\in b$ given a set of plaintext states $\set{P}$ there
exists a PQC sending the set $\set{P}$ to $\op{\rho}^{(0)}$. In
Section \ref{sec:Realiza-entropy} we will analyze concrete PQC realizations as well as the minimal entropy of the key for a given set $\set{P}$ and $\op{\rho}^{(0)}$. We will show an easily understandable geometrical method how to construct PQCs for encryption of the given set of plaintexts.

%%%%%%%%%%%%%%%%%%%%%%%%%%%%%%%%%%%%%%%%%%%%%%%%%%%%%%%%%%%%%%%%%%%%%%%%%%%%%%%%%%%%%%%%%%%%%%%%%%%%%%
\section{General remarks}
\label{sub:General-analysi}
%%%%%%%%%%%%%%%%%%%%%%%%%%%%%%%%%%%%%%%%%%%%%%%%%%%%%%%%%%%%%%%%%%%%%%%%%%%%%%%%%%%%%%%%%%%%%%%%%%%%%%

In our analysis we will exploit the geometric picture of the Bloch ball (see Appendix). In both cases we will
define specific representatives of the set of plaintexts $\overline{\set{P}}$. We will choose the basis
of the state space as four operators $\op{\xi}_j$ represented by
mutually orthogonal Bloch vectors $\vec{v}_j$. In particular, we will
rotate the coordinate system (this rotation is just unitary change
of the basis operators) to work with not necessarily positive,
but trace-one operators
\begin{equation}
\begin{array}{rclcl}
\op{\xi}_x &=&  \frac{1}{2}(\id+ \alpha \op{S}_x)  &\leftrightarrow & \vec{v}_x=(\alpha,0,0)\\
\op{\xi}_y &=&  \frac{1}{2}(\id+ \beta \op{S}_y) &\leftrightarrow & \vec{v}_y=(0,\beta,0) \\
\op{\xi}_z &=&  \frac{1}{2}(\id+ \op{S}_z)  &\leftrightarrow & \vec{v}_z=(0,0,1)\\
\op{\xi}_0 &=&  \frac{1}{2}\id  &\leftrightarrow & \vec{v}_0=(0,0,0)\\
\end{array}
\label{ksi}
\end{equation}
The S-basis is just a suitably rotated $\sigma$-basis (basis consisting of Pauli operators),
i.e. $\op{S}_j=\op{U}\op{\sigma}_j \du{\op{U}}$ for some unitary $\op{U}$.

This new operator basis shares all the properties of the original
Pauli basis. In fact, the operators $\op{S}_x,\op{S}_y,\op{S}_z$ specify only
a rotated Cartesian coordinate system. Each private quantum channel
$\sop{E}$ induces a contraction of the given set $\overline{\set{P}}$
into the state $\op{\rho}^{(0)}$. Our first aim is to explicitly specify the maximally
mixed state in $\overline{\rho}\in\overline{\set{P}}$. The second
goal will be to show the achievability of this state, i.e.
the construction of the PQC that transforms the whole set
of plaintexts states into the state $\op{\rho}^{(0)}$ having the same mixedness (i.e. distance from $\frac{1}{2}\id$)
as $\overline{\rho}$. In particular,
$\op{\rho}^{(0)}=\op{V}\overline{\rho}\du{\op{V}}$ ($\op{V}$ is unitary).
Let us denote by $s$ the mixedness of $\overline{\rho}$. Then the
PQC $\sop{E}_{\op{\rho}^{(0)}}$ acts (in a suitably chosen basis) as follows:
$\sop{E}_{\op{\rho}^{(0)}}[\op{\xi}_j]=\op{\rho}^{(0)}=\frac{1}{2}(\id+s\cdot \op{S}_z)$ for
all $\op{\xi}_j\in\overline{\set{P}}$. Its action on the linear complement of $\overline{\set{P}}$ must be defined in a way that the whole transformation is TCP. The existence of such PQC will be proved in subsequent sections.

%%%%%%%%%%%%%%%%%%%%%%%%%%%%%%%%%%%%%%%%%%%%%%%%%%%%%%%%%%%%%%%%%%%%%%%%%%%%%%%%%%%%%%%%%%%%%%%%%%%%%%%%%%%%%%%%%%%%%%%%
\section{Two states}
\label{sub:two_states}
%%%%%%%%%%%%%%%%%%%%%%%%%%%%%%%%%%%%%%%%%%%%%%%%%%%%%%%%%%%%%%%%%%%%%%%%%%%%%%%%%%%%%%%%%%%%%%%%%%%%%%%%%%%%%%%%%%%%%%%%

Given two linearly independent states $\set{P}=\{\op{\rho}_1,\op{\rho}_2\}$ the set $\overline{\set{P}}$ defines a line crossing the Bloch sphere in two pure states, e.g.  $|\psi_1\rangle,|\psi_2\rangle$. The mixedness of $\op{\rho}_\lambda=\lambda\op{\rho}_1+(1-\lambda)\op{\rho}_2$ (i.e. the distance from the total mixture) is characterized by the length of the corresponding Bloch vector $\vec{r}_\lambda$. In particular, $|\vec{r}_\lambda|^2=\lambda^2 |\vec{r}_1|^2+(1-\lambda)^2 |\vec{r}_2|^2+2\lambda(1-\lambda)\vec{r}_1\cdot\vec{r}_2$ can be easily minimized (with respect to $\lambda$) providing that we use two pure states $|\psi_j\rangle\leftrightarrow\vec{r}_j$. In this case $|\vec{r}_1|=|\vec{r}_2|=1$ and
$\vec{r}_1\cdot\vec{r}_2=|\vec{r}_1|\cdot|\vec{r_2}|\cos\theta$ with $\theta\in[0,\pi]$ being an angle between the vectors. The minimum we obtain by calculating the equation
\begin{equation}
\frac{d}{d\lambda}[\lambda^2+(1-\lambda)^2-2\lambda(1-\lambda)\cos\theta]=
2(2\lambda-1)(1-\cos\theta)=0
\end{equation}
The minimum is achieved for $\lambda=1/2$, i.e. for the equal mixture of two pure states from $\overline{\set{P}}$ and reads $|\vec{r}_{\rm min}|=\sqrt{\frac{1}{2}(1+\cos\theta)}$. For general (nonpure) states $\op{\rho}_1\leftrightarrow\vec{r}_1$ and $\op{\rho}_2\leftrightarrow\vec{r}_2$ the state $\op{\rho}_\lambda=\lambda\op{\rho}_1+(1-\lambda)\op{\rho}_2$ is maximally mixed for the value
$\lambda=(|\vec{r}_2|^2-\vec{r}_1\cdot\vec{r}_2)/|\vec{r}_1-\vec{r}_2|^2$.

Let us assume that the state $\ket{\psi_1}$ corresponds to the North Pole of the Bloch ball and the $y$ coordinate of $\ket{\psi_2}$ vanishes, i.e. we choose the operator basis $\op{S}_x,\op{S}_y,\op{S}_z$  such that
$\ket{\psi_1}\bra{\psi_1}=\frac{1}{2}(\id+\op{S}_z)$ and
$\ket{\psi_2}\bra{\psi_2}=\frac{1}{2}(\id+\sin\theta \op{S}_x+\cos\theta
\op{S}_z)$. In other words, the state $\ket{\psi_2}$ is represented by the
vector $\vec{r}_2=(\sin\theta,0,\cos\theta)$. The norm of the vector
$\vec{r}_\lambda$ is minimal for $\lambda_{\rm min}=1/2$,
i.e. $\vec{r}_{\rm min}=(\frac{1}{2}\sin\theta,0,\frac{1}{2}(1+\cos\theta))$
with norm $\abs{\vec{r}_{\rm min}}=\sqrt{\frac{1}{2}(1+\cos\theta)}$.

The possible quantum private channels form a set
\begin{equation}
\sop{E}=\left(\ba{cccc}
1 & 0 & 0 & 0 \\
0 & \frac{1}{2}(1-\cos\theta) & a & \frac{1}{2}\sin\theta \\
0 & 0 & b & 0 \\
0 & \frac{1}{2}\sin\theta & c & \frac{1}{2}(1+\cos\theta) \\
\ea\right)
\end{equation}
and the complete positivity with respect to parameters $a$,
$b$ and $c$ must be verified. Our aim is to find at least one
valid TCP transformation. Therefore, let us consider that only the parameter
$b$ is nonvanishing\footnote{We will show that even in this particular case there exists a completely positive superoperator.} (i.e. we set $a=c=0$). In this case the
matrix is symmetric, so the singular values
coincide with the eigenvalues that read
\begin{equation}
\{\lambda_1,\lambda_2,\lambda_3\}=\{1,b,0\}.
\end{equation}

The complete positivity constraint
\cite{Fujiwara+Algoet-One-t_param_quant:1999}
requires the validity of the following inequalities
\begin{eqnarray}
\label{in1}
1+\lambda_1-\lambda_2-\lambda_3 \ge 0 & \Rightarrow &
b\le 2\\
\label{in2}
1-\lambda_1+\lambda_2-\lambda_3\ge 0 & \Rightarrow &
b\ge 0\\
\label{in3}
1-\lambda_1-\lambda_2+\lambda_3\ge 0 & \Rightarrow &
b\le 0 \\
\label{in4}
1+\lambda_1+\lambda_2+\lambda_3\ge 0 & \Rightarrow &
b \ge -2.
\end{eqnarray}
It turns out that the only possibility to satisfy these
conditions is that the value of $b$ must set to zero, i.e. $b=0$.
As a result, we get that the channel $\sop{E}$ with $b=0$ is for
sure completely positive for all values of $\cos\theta$. Consequently,
for two linearly independent states the derived bound on the choice
of the state $\op{\rho}^{(0)}$ is achievable. The achievability of the states inside the ball $b$ we obtain from the fact that the set of PQCs encrypting given set $\set{P}$ is convex as well as the set of all achievable states, see Section \ref{sec:Achievable-states}. Later we will specify the unitary
transformations forming the private quantum channel explicitly.

%%%%%%%%%%%%%%%%%%%%%%%%%%%%%%%%%%%%%%%%%%%%%%%%%%%%%%%%%%%%%%%%%%%%%%%%%%%%%%%%%%%%%%%%%%%%%%%%%%%%%%%%%%%%%%%%%%%%%%%%%%
\section{Three linearly independent states}
\label{sub:Three-linearl-indepen-states}
%%%%%%%%%%%%%%%%%%%%%%%%%%%%%%%%%%%%%%%%%%%%%%%%%%%%%%%%%%%%%%%%%%%%%%%%%%%%%%%%%%%%%%%%%%%%%%%%%%%%%%%%%%%%%%%%%%%%%%%%%%

In case the set $\set{P}=\{\op{\rho}_1,\op{\rho}_2,\op{\rho}_3\}$ contains
precisely three linearly independent states, the set
$\overline{\set{P}}$ forms a
plane and valid quantum states from this plane (the intersection with
the Bloch ball) form a circle $c$. Since all points of the ball $b$,
containing all possible candidates for the state $\op{\rho}^{(0)}$, have the
distance from $\frac{1}{2}\id$ the same or smaller than the most mixed state
from $c$, it follows that the circle $c$ touches the ball $b$
precisely in the middle of the circle $c$. Moreover,
this point $\overline{\rho}=\op{\rho}^{(0)}$ is the most mixed state from $c$.

To solve the general case explicitely, we will exploit the tools of analytic
geometry. A plane determined by three points $A=\vec{r}_1$,
$B=\vec{r}_2$, $C=\vec{r}_3$ reads $ax+by+cz+d=0$, where
\begin{eqnarray}
d&=&\det (\vec{r}_1\ \vec{r}_2\ \vec{r}_3) \\
a&=&\det (\vec{1}\ \vec{r}_2\ \vec{r}_3) \\
b&=&\det (\vec{r}_1\ \vec{1}\ \vec{r}_3) \\
c&=&\det (\vec{r}_1\ \vec{r}_2\ \vec{1}).
\end{eqnarray}
The symbol  $\vec{1}=(1,1,1)^T$ denotes a column vector. The distance
from the origin of the coordinate system (the total mixture) equals to
\begin{equation}
s=\frac{|d|}{\sqrt{a^2+b^2+c^2}}.
\end{equation}
This number coincides (if $s\le 1$) with the distance between the maximally mixed state $\overline{\rho}\in\overline{\set{P}}$ and the total mixture. It follows that we can use directly the given set of plaintext states $\{\op{\rho}_1,\op{\rho}_2,\op{\rho}_3\}$ as the basis. However, for our purposes
it will be useful to choose operators $\op{\xi}_1,\op{\xi}_2,\op{\xi}_3\in\overline{\set{P}}$
of the form given in Eq. (\ref{ksi}). Let us assume that the set $\overline{\set{P}}$ does not contain the total mixture. In this case
\begin{equation}
a=\beta\ \ \ \
b=\alpha\ \ \ \
c=d=\alpha\beta
\end{equation}
and
\begin{equation}
\label{s}
s=|\vec{s}|=\frac{|\alpha\beta|}{\sqrt{\alpha^2\beta^2+\beta^2+\alpha^2}}.
\end{equation}
The question is, whether the transformations
$\op{\xi}_j\mapsto\op{\rho}^{(0)}=\frac{1}{2}(\id+s\cdot \op{S}_z)=\sop{E}[\op{\xi}_j]$
is completely positive, or not. Due to the unitality of the PQC
channels, the transformation $\sop{E}$ is completely specified as
\begin{equation}
\sop{E}=\left(
\begin{array}{cccc}
1 & 0 & 0 & 0 \\
0 & 0 & 0 & 0 \\
0 & 0 & 0 & 0 \\
0 & s/\alpha & s/\beta & s \\
\end{array}\right).
\end{equation}
Our task is only to verify the condition of complete positivity.
The singular values of $\sop{E}$ reads
$\{\lambda_1,\lambda_2,\lambda_3\}=
\{0,0,s\sqrt{1+\frac{1}{\alpha^2}+\frac{1}{\beta^2}}\}$.
It follows that the map is completely positive if and only if
$0\le 1-s\sqrt{1+\frac{1}{\alpha^2}+\frac{1}{\beta^2}}$.
Inserting the derived result for the value of $s$
into this complete positivity constraint,
we find that it is always satisfied, because
\begin{equation}
0\le 1-s\sqrt{1+\frac{1}{\alpha^2}+\frac{1}{\beta^2}}=
1-\frac{|\beta\alpha|}{\sqrt{\beta^2+\alpha^2+\beta^2\alpha^2}}
\sqrt{\frac{\beta^2\alpha^2+\alpha^2+\beta^2}{\beta^2\alpha^2}}
=1-1=0.
\end{equation}

As a result we obtain that it is always possible to define private quantum
channel so that the norm bound is saturated and the state
$\op{\rho}^{(0)}=\overline{\rho}$ is achievable. Let us note that
the value of $s$ is always less than 1, which is in agreement with the
fact that $\op{\rho}^{(0)}$ is a quantum state (it belongs to the Bloch
ball). The fact that for a given set of states $\set{P}$ it is always
possible to find a unital channel $\sop{E}$ such that $\forall\
\op{\rho}\in\set{P}:\sop{E}(\op{\rho})=\op{\rho}^{(0)}$, where $\op{\rho}^{(0)}$ is the closest
state to the total mixture belonging to the linear span of the set $\set{P}$,
is interesting {\it per se}.

The case when $\op{\rho}^{(0)}$ lies inside (not on the surface) the ball $b$ we obtain again from the convexity of the set of private quantum channels as discussed in Section \ref{sec:Achievable-states}.

%%%%%%%%%%%%%%%%%%%%%%%%%%%%%%%%%%%%%%%%%%%%%%%%%%%%%%%%%%%%%%%%
\section{Realizations of PQC and entropy of the key}
\label{sec:Realiza-entropy}
%%%%%%%%%%%%%%%%%%%%%%%%%%%%%%%%%%%%%%%%%%%%%%%%%%%%%%%%%%%%%%%%

So far, we studied the existence of PQC with states $\op{\rho}^{(0)}$ for a
given set of plaintexts $\overline{\set{P}}$. Next we will analyze
the optimal realizations of these private quantum channels, i.e. we
will ask the question: how many classical bits one needs to design
the PQC. These classical bits represent the key that must be shared
between sender and receiver to perfectly encrypt/decrypt the quantum
states from plaintext. The efficiency of PQC is quantified by the
entropy ($H(p)=-\sum_j p_j\log_2 p_j$)
of the probability distribution of unitary transformations that
specify the length of the shared classical key. It is known
\cite{Ambainis+Mosca...-Priva_quant_chann:2000}
that the ideal realization of PQC for general qubit states
requires two bits. It is realized by arbitrary collection of
four unitary transformations $\{\op{U}_k\}_k$ satisfying the following
orthogonality condition ${\rm Tr} (\du{\op{U}_j} \op{U}_k)=2\delta_{jk}$.
Each of these transformations is applied with the same probability
$p=1/4$, i.e. $H(p)=2$.

Each private quantum channel $\sop{E}$ is a convex combination
of unitary transformations. Our task is to find a representation
for arbitrary PQC, which is optimal. The action channel $\sop{E}$ can be
written in the form
$\sop{E}[\op{\rho}]=\op{U}\Phi_{\sop{E}}[\op{V}\op{\rho} \du{\op{V}}]\du{\op{U}}$,
where $\op{U},\op{V}$ are unitary transformations. For qubit unital channels the
induced transformation $\Phi_{\sop{E}}$ is diagonal, i.e.
$\Phi_\sop{E}={\rm diag}\{1,\lambda_1,\lambda_2,\lambda_3\}$. It turns
out that these transformations are of a simple form and can be written
as convex combination of four Pauli transformations
\begin{equation}
\Phi_\sop{E}[\op{\rho}]=p_0\op{\rho}+p_x\op{\sigma}_x\op{\rho}\op{\sigma}_x
+p_y\op{\sigma}_y\op{\rho}\op{\sigma}_y+p_z\op{\sigma}_z\op{\rho}\op{\sigma}_z
\end{equation}
Consequently, the original transformation $\sop{E}$ is realized
by four unitary transformations $\op{W}_j=\op{U}\op{\sigma}_j \op{V}$, i.e.
$\sop{E}[\op{\rho}]=\sum_j p_j \op{W}_j\op{\rho}\du{\op{W}}_j$.
Since the probabilities do not change,
the PQCs $\sop{E}$ and $\Phi_\sop{E}$ can be realized
with the same entropy. In fact, this holds in general: two unitarily
equivalent PQCs can be always realized with the same efficiency, i.e.
with the classical keys of the same entropy. Thus, it is sufficient to
analyze the optimality of the realization of Pauli channels
$\Phi_\sop{E}$.
Finding the singular values corresponding to $\sop{E}$ we obtain the
diagonal elements $\lambda_1,\lambda_2,\lambda_3$ of $\Phi_\sop{E}$.
The probabilities $p_j$ are related to these values $\lambda_k$
via the following equations
\begin{equation}
\begin{array}{rcl}
p_x&=& \frac{1}{4}(1+\lambda_1-\lambda_2-\lambda_3) \\
p_y&=& \frac{1}{4}(1-\lambda_1+\lambda_2-\lambda_3) \\
p_z&=& \frac{1}{4}(1-\lambda_1-\lambda_2+\lambda_3) \\
p_0&=& 1-p_x-p_y-p_z\\
\end{array}
\end{equation}

The entropy rate of the given PQC $H(\sop{E})=H(p)$, $p=\{p_j\}_j$, is given by
the entropy of the distribution $p$. Let $\op{\rho}=\sum_j p_j\ket{\psi_j}\bra{\psi_j}$, where $\{\ket{\psi_j}\}_j$ is a set of not necessarily orthogonal quantum states. It follows that
$S(\op{\rho})=S(\sum_j p_j\ket{\psi_j}\bra{\psi_j})\le H(p)$ and the inequality
is saturated if and only if $\{\ket{\psi_j}\}_j$ are mutually orthogonal. Let us consider any pure plaintext state $\ket{\psi}$ and let $\ket{\psi_j}=\op{U}_j\ket{\psi}$. It is clear that
\begin{equation}
\label{equ:entropy_limit_below}
S(\sop{E}(\ket{\psi}\bra{\psi}))=S(\op{\rho})=S\left(\op{\rho}^{(0)}\right)\le H(p).
\end{equation}
Therefore the entropy of the encryption operation can always be bounded from below by the entropy of $\op{\rho}^{(0)}$ as long as $\overline{\set{P}}$ contains at least one pure state. This always holds in the case of qubit, however, not in general for systems of larger dimension.
In example in the case of two qubits we can define the set $\set{P}=\{1/4\id,1/2(\ket{00}\bra{00}+\ket{11}\bra{11})\}$. It is clear that $\overline{\set{P}}$ contains no pure state, it is encrypted by the superoperator
\begin{equation}
\{(1/2,\id),(1/2,\id\otimes\op{\sigma}_x)\}
\end{equation}
and $\op{\rho}^{(0)}=\frac{1}{4}\id$.

The limit \eqref{equ:entropy_limit_below} is saturated
only if the encoding operators $\op{U}_j$ generate mutually orthogonal
(noncommuting) states (ciphertexts) for each given plaintext. In
particular, if we consider a PQC for all possible states of a qubit, this limit
can be achieved only (up to unitary equivalence) by encoding with the identity and the {\it universal NOT} operation.
However, this map is not completely positive, and therefore unphysical
\cite{Buzek+Hilery...-Optim_manip_qubit:1999}.

Next we shall study the realization of PQC when the set of
plaintexts is two-dimensional and three-dimensional, respectively, and
the state $\op{\rho}^{(0)}$ is the maximally mixed one from the set $\overline{\set{P}}$.
For two-dimensional set of plaintexts the induced transformation is
$\Phi_\sop{E}={\rm diag}\{1,0,0,1\}$, i.e. $\lambda_1=\lambda_2=0$
and $\lambda_3=1$. It follows that
\begin{equation}
\Phi_\sop{E}[\op{\rho}]= \frac{1}{2}\op{\rho}+\frac{1}{2}\op{\sigma}_z\op{\rho}\op{\sigma}_z
\end{equation}
and one bit is sufficient for encoding.

One can specify the precise form of unitary transformations, but the explicit calculation is quite
lengthy. Instead, we will use the geometric picture of Bloch sphere
to guess the unitaries.

%%%%%%%%%%%%%%%%%%%%%%%%%%%%%%%%%%%%%%%%%%%%%%%%%%%%%%%%%%%%%%%%%%%%%%%%%%%%%%%%%%%%%%%%%%%%%%%%%%%%%%%%%%%%%%%%
\subsection*{Two states}

\begin{figure}
  \begin{center}
  \includegraphics[width=7cm]{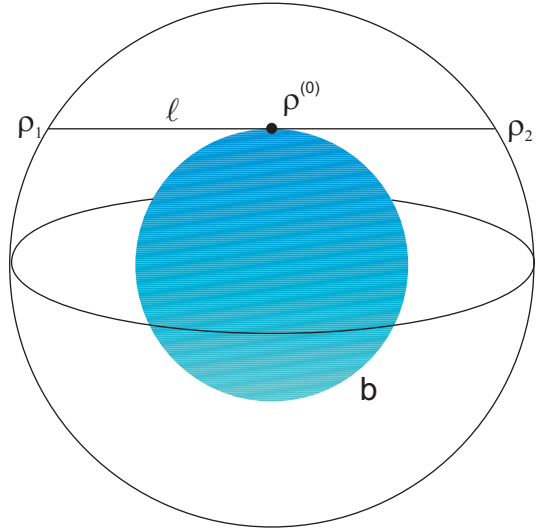}
  \caption{Encrypting line segment $l$.}
  \label{PQC3d1}
  \end{center}
\end{figure}

Let us suppose that the set $\set{P}=\{\op{\rho}_1,\op{\rho}_2\}$ has only two (linearly independent) states. By Theorem \ref{the:PQC_linear_span} it also encrypts any state on the line segment $l\subseteq\overline{\set{P}}$ defined by the points corresponding to the states $\op{\rho}_1$ and $\op{\rho}_2$ in the Bloch ball.
Let us choose the state $\op{\rho}^{(0)}$ as the point where the line segment $l$ touches the ball\footnote{Of possible candidates to the state $\op{\rho}^{(0)}$.} $b$ (see fig. \ref{PQC3d1}), i.e. $\op{\rho}^{(0)}=\overline{\rho}$. It is clear that this point is in the center of the line segment $l$, since the extremal points of this line segment are on the surface of the Bloch ball.

We will design a specific superoperator $\sop{E}$, which encrypts this line segment to the state $\op{\rho}^{(0)}$. This superoperator can be realized using two unitary operators with uniform distribution,
\begin{equation}
\label{equ:PQC_two_states_sop}
\sop{E}(\op{\rho})=\frac{1}{2}\id\op{\rho}\id+\frac{1}{2}\op{U}\op{\rho}\du{\op{U}},
\end{equation}
where $\op{U}$ is the unitary operation, which realizes the rotation of the Bloch ball by $180$ degrees around the axis intersecting points $\op{\rho}^{(0)}$ and $\frac{1}{2}\id$. It is easy to see that such a superoperator takes any state $\op{\rho}$ from $l$ to a convex combination of the original state $\op{\rho}$ and the state which lies on the line $l$ in the same distance from $\op{\rho}^{(0)}$, but on the opposite half line (starting in the point $\op{\rho}^{(0)}$). The consequence is that the convex combination
\begin{equation}
\frac{1}{2}\op{\rho}+\frac{1}{2}\op{U}\op{\rho}\du{\op{U}}=\op{\rho}^{(0)}.
\end{equation}
The way to achieve any other point lying on the surface of the ball $b$ is straightforward. For any such point $\op{\rho}^{\prime(0)}$ there exists a two dimensional rotation $\op{R}_{\op{\rho}^{(0)},\op{\rho}^{\prime(0)}}$, which rotates the point $\op{\rho}^{(0)}$ to the point $\op{\rho}^{\prime(0)}$. This rotation is realized by some unitary operation $\op{U}_{\op{\rho}^{(0)},\op{\rho}^{\prime(0)}}$ on the density operators. Therefore, the superoperator encrypting the whole line segment $l$ into the point $\op{\rho}^{\prime(0)}$ is
\begin{equation}
\label{equ:PQC_two_states_sop_rotated}
\sop{E}(\op{\rho})=\frac{1}{2}\op{U}_{\op{\rho}^{(0)},\op{\rho}^{\prime(0)}}\id\op{\rho}\id\du{\op{U}}_{\op{\rho}^{(0)},\op{\rho}^{\prime(0)}}
+\frac{1}{2}\op{U}_{\op{\rho}^{(0)},\op{\rho}^{\prime(0)}}\op{U}\op{\rho}\du{\op{U}}\du{\op{U}}_{\op{\rho}^{(0)},\op{\rho}^{\prime(0)}}.
\end{equation}
This superoperator has again Kraus decomposition with only two unitary operators, and therefore only a single bit of key is needed. The method how to achieve any $\op{\rho}^{\prime(0)}\in b$ (not only on the surface) is based on the method of encryption of three linearly independent states and will be discussed later in this section.

In this way we have demonstrated that one bit of key is sufficient to
encrypt set $\set{P}$ containing two linearly independent states. It
remains to verify whether one bit is also necessary.
There might e.g. exist some encryption operation
$\{(p_1,\op{U}_1),(p_2,\op{U}_2)\}, p_1+p_2=1, p_1\ne p_2$ encrypting the
set $\set{P}$. Clearly the entropy of the key of such an operation is
smaller than one. We will prove that one bit is necessary by showing
that any encryption superoperator $\sop{E}$ encrypting a line $l$
encrypts also the line $l'$, which is parallel to $l$ and intersects
$\frac{1}{2}\id$. Then the derived inequality \eqref{equ:entropy_limit_below}
implies that one bit is indeed necessary, since $S(\frac{1}{2}\id)=1$.
Also, $\sop{E}$ can be used to encrypt a classical bit ($\{\ket{0},\ket{1}\}$) and this result
is in accordance with \cite{Shannon-CommunicationTheoryof-1949}.

\begin{figure}
  \begin{center}
  \includegraphics[width=7cm]{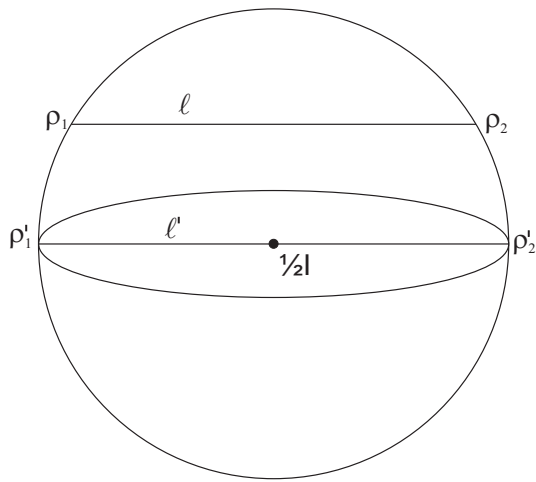}
  \caption{Encryption of $l$ encrypts also $l'$.}
  \label{PQC3d2}
  \end{center}
\end{figure}

Let us denote $\op{\rho}_1$ and $\op{\rho}_2$ the extremal points of the line segment $l$ (lying on the surface of the Bloch ball, see figure \ref{PQC3d2}) and $\op{\rho}_1'$ and $\op{\rho}_2'$ the extremal points of the line segment $l'$. Let us express each of the points $\op{\rho}_1'$ and $\op{\rho}_2'$ as a linear combination of the points $\op{\rho}_1$, $\op{\rho}_2$ and $\frac{1}{2}\id$. It is easy to see that the coordinates of the points satisfy
\begin{equation}
\op{\rho}_1'=x\op{\rho}_1+y\op{\rho}_2+z\frac{1}{2}\id\ \ \Leftrightarrow\ \ \op{\rho}_2'=y\op{\rho}_1+x\op{\rho}_2+z\frac{1}{2}\id.
\end{equation}
This relation holds for extremal points of any line segment parallel to $l$ and lying in the plane spanned by $l$ and $\frac{1}{2}\id$. Let $\sop{E}$ encrypts $l$ to some state $\op{\rho}^{(0)}$. Then from linearity and unitality of $\sop{E}$ we obtain
\begin{equation}
\label{equ:encryption_l-l}
\sop{E}(\op{\rho}_1')=x\sop{E}(\op{\rho}_1)+y\sop{E}(\op{\rho}_2)+z\frac{1}{2}\id= y\sop{E}(\op{\rho}_1)+x\sop{E}(\op{\rho}_2)+z\frac{1}{2}\id=\sop{E}(\op{\rho}_2')
\end{equation}
since $\sop{E}(\op{\rho}_1)=\sop{E}(\op{\rho}_2)=\op{\rho}^{(0)}$ from assumption that $\sop{E}$ encrypts $l$. It follows that $\sop{E}$ encrypts any line segment parallel to $l$ lying in the plane spanned by $l$ and $\frac{1}{2}\id$.

From Eq. \eqref{equ:encryption_l-l} we can also easily determine the state $\op{\rho}'^{(0)}=\sop{E}(\op{\rho}_1')$, i.e. the state where $\sop{E}$ sends the line segment $l'$. It is the state $\op{\rho}^{(0)}$ shifted towards $\frac{1}{2}\id$, from the equation
\begin{equation}
\sop{E}(\op{\rho}_1')=x\sop{E}(\op{\rho}_1)+y\sop{E}(\op{\rho}_2)+z\frac{1}{2}\id=(x+y)\op{\rho}^{(0)}+z\frac{1}{2}\id.
\end{equation}
The ratio between $(x+y)$ and $z$ determines the distance from $\frac{1}{2}\id$.

We can also derive an analogical result for PQC encrypting a circle in the Bloch sphere. In this case the fact that it also encrypts all parallel circles implies that this PQC establishes an approximative encryption of the whole Bloch sphere, as defined in \cite{Hayden+Leung...-Rando_quant_state:2003}. Also, in general, PQC encrypting any set $\overline{\set{P}}$ in any Hilbert space encrypts also all spaces parallel in the superplane spanned by $\overline{\set{P}}$ and $\frac{1}{d}\id$. This will be discussed in detail in a separate paper.

%%%%%%%%%%%%%%%%%%%%%%%%%%%%%%%%%%%%%%%%%%%%%%%%%%%%%%%%%%%%%%%%%%%%%%%%%%%%%%%%%%%%%%%%%%%%%%%%%%%%%%%%%%%%%%%%%%%%%%%%%%
\subsection*{Three states}

\begin{figure}
  \begin{center}
  \includegraphics[width=7cm]{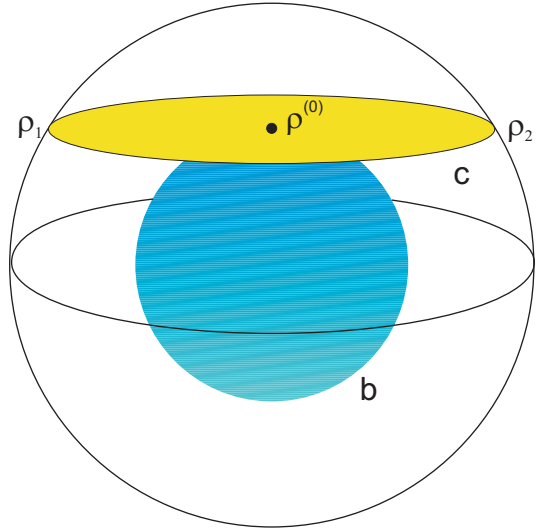}
  \caption{Encryption of the circle $c$.}
  \label{PQC3d3}
  \end{center}
\end{figure}

In case the set $\set{P}$ contains precisely three linearly
independent states $\op{\rho}_1,\op{\rho}_2,\op{\rho}_3$, their linear span is a plane and valid quantum states from this plane (the intersection with the Bloch ball) form a circle $c$ (see figure \ref{PQC3d3}). Since all points of the ball $b$, containing all possible candidates for the state $\op{\rho}^{(0)}$, have the distance from $\frac{1}{2}\id$ the same or smaller than the most mixed state from $c$, it follows that the circle $c$ touches the ball $b$ precisely in the middle of the circle $c$. Moreover, this point $\overline{\rho}=\op{\rho}^{(0)}$ is the most mixed state from $c$.

Following analogical argumentation as in the case of the two states, we construct the TCP superoperator, which encrypts the whole circle and sends it to $\op{\rho}^{(0)}$. This superoperator is the same as in the case of two states, it is the superoperator \eqref{equ:PQC_two_states_sop}. The operator $\op{U}$ is the rotation around the axis of the circle $c$. The saturation of any other point on the surface of the ball $b$ is the same as in the case of two states, see Eq. \eqref{equ:PQC_two_states_sop_rotated}.

%This is quite interesting, because the operator $\op{U}$ in equation \eqref{equ:PQC_two_states_sop} is the universal NOT on the circle $c$ and such an universal NOT can be designed for any circle inside the Bloch Ball. Thus, the impossibility of the universal NOT operation of the whole Bloch ball \cite{Buzek+Hilery...-Optim_manip_qubit:1999,Buzek+Hilery...-Unive_gate:2000} can be derived from the fact that two bits are necessary to encrypt a qubit.

\begin{figure}
  \begin{center}
  \includegraphics[width=7cm]{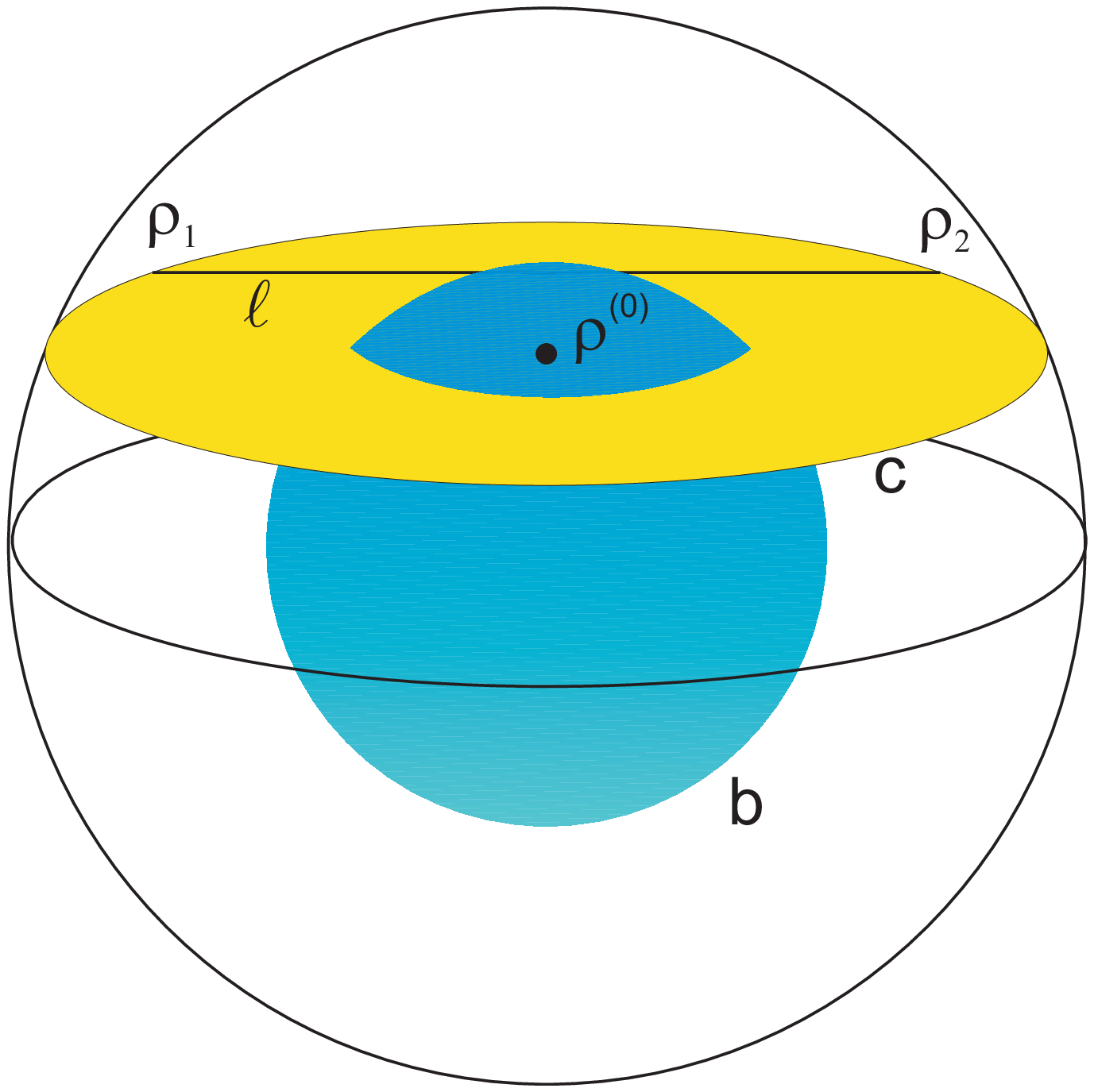}
  \caption{Encryption of the line $l$ with $D(\op{\rho}^{(0)},1/2\id)<D(\overline{\rho},1/2\id)$.}
  \label{PQC3d4}
  \end{center}
\end{figure}

Using this result we may also design a PQC, which encrypts a set
$\set{P}$ of two linearly independent states into the arbitrary state
$\op{\rho}^{(0)}$ inside the ball $b$ by using just single bit of
key. The ball $b$ is now specified by the given line $l$ associated
with the set $\overline{\set{P}}$.
The state $\op{\rho}^{(0)}$ specifies uniquelly a sphere $g$ of the radius
$r=D(\op{\rho}^{(0)},\frac{1}{2}\id)$, centered
in total mixture and containing this state on its surface. There
exists a tangent plane $\kappa$ to this sphere determined by the original
line $l$. This plane is generated by three linearly independent
states and following the reasoning of this subsection, it can be
encrypted  into its maximally mixed state by PQC with $H(\sop{E})=1$.
However, the maximally mixed state equals to the only point in the
intersection of the plane with the sphere (see figure \ref{PQC3d4}).
This point is unitarily equivalent to $\op{\rho}^{(0)}$. It means that
the original set
$\overline{\set{P}}$ given by two linearly independent states (forming
the line in the plane, $l\in\kappa$) is encrypted by the same PQC (up
to unitary transformation) into
the state $\op{\rho}^{(0)}$.

Let us now proceed with the analytic approach to see how large key is required to encrypt the set $\overline{\set{P}}$ into arbitrary state $\op{\rho}^{(0)}\in b$. For three-dimensional set of plaintexts we
have a unique PQC that transforms $\overline{\set{P}}$ into the state $\op{\rho}^{(0)}$. The singular values of the corresponding mapping $\sop{E}$ reads $\{\lambda_1,\lambda_2,\lambda_3\}
=\{0,0,s\sqrt{1+\frac{1}{\alpha^2}+\frac{1}{\beta^2}}\}$. It follows that
\begin{equation}
\label{equ:Pauli_chan_probab}
\begin{array}{lcr}
p_0=p_z&=&(1+s\sqrt{1+\frac{1}{\alpha^2}+\frac{1}{\beta^2}})/4 \\
p_x=p_y&=&(1-s\sqrt{1+\frac{1}{\alpha^2}+\frac{1}{\beta^2}})/4 \\
\end{array}
\end{equation}
The parameter $s$ corresponds to the distance between the
state $\op{\rho}^{(0)}$ and the total mixture. It is bounded by the inequality
$s\le\frac{|\alpha\beta|}{\sqrt{\alpha^2\beta^2+\alpha^2+\beta^2}}$.
Except the case when this inequality is saturated we need four unitary
transformations to realize the PQC. For maximal value of $s$ two
unitary transformations are sufficient. Moreover, they are used with
equal probabilities. It means that the limiting case
($\op{\rho}^{(0)}=\overline{\rho}$) for two-dimensional and three-dimensional
set of plaintexts has the same entropy rates.

If you examine the dependence of the probabilities $\{p_i\}_i$ on the parameter $s$, then you realize that to encrypt three states a single bit of key is sufficient if $\op{\rho}^{(0)}$ is on the surface of the ball $b$ (see the beginning of this section). However, as the state $\op{\rho}^{(0)}$ is getting closer to $\frac{1}{2}\id$, the entropy of the key grows up to two bits.

The key question is whether the derived bound on entropy of PQC encrypting the given set of three linearly independent states is also necessary, i.e. whether it is possible for given $\set{P}$ and $\op{\rho}^{(0)}$ design a PQC with lower entropy of key than in the derived example. Let us recall the proof that one bit of key is necessary to encrypt the set of two linearly independent plaintexts, see Section \ref{sub:two_states}. It immediately follows that one bit is also necessary to encrypt plaintext containing at least three linearly independent states.

The final step is to prove that also the entropy of the private quantum channel encrypting three dimensional set $\set{P}$, with $D(\op{\rho}^{(0)},\frac{1}{2}\id)<D(\overline{\rho},\frac{1}{2}\id)$, realized using the Pauli channel is minimal. Let us introduce the entropy exchange $S_{ex}(\op{\rho}_A,\sop{E})$ \cite{Nielsen+Chuang-Quant_Compu_Quant:2000} as the quantity measuring the part of quantum information which is lost into the environment under the action of the channel $\sop{E}$ providing that the system is initially prepared in the state $\op{\rho}_A$. Provided that the channel $\sop{E}$ on the system $A$ is realized using unitary operation $\op{G}$ on a larger system $AE$, where $E$ is the environment, the entropy of exchange is defined as the von Neumann entropy of the reduced density matrix of the environment after applying the operation $\op{G}$. It turns out that this entropy is independent of concrete realization of the superoperator.

In particular, any channel $\sop{E}$ has unitary representation $\sop{E}[\op{\rho}_A]={\rm Tr}_E[\op{G}(\op{\rho}_A\otimes\ket{0}\bra{0})\du{\op{G}}]=\sum_j \op{A}_j\op{\rho}_A \du{\op{A}}_j$ with
$\op{G}=\sum_j \op{A}_j\otimes \ket{j}\bra{0}$. It is known that the entropy of the environment state
$\op{\omega}_E={\rm Tr}_A[\op{G}(\op{\rho}_A\otimes\ket{0}\bra{0})\du{\op{G}}]
=\sum_{jk} {\rm Tr}[\op{A}_j\op{\rho}_A \du{\op{A}}_k] \ket{j}\bra{k}$ does not
depend on the particular Kraus representation.

In our case this function is the lower bound of the entropy of the key
$H(\{p_k\}_k)$, i.e. $H(\{p_k\}_k)\ge \max_{\op{\rho}_A} S_{ex}(\op{\rho},\sop{E})$. This inequality follows from the fact that $S(\op{\omega}_E)\le S({\rm diag}_{\mathcal B}[\op{\omega}_E])$\footnote{The ${\rm diag}_{\mathcal B}[\op{\omega}_E]$ is the all-zero matrix except for the diagonal elements, which are equal to diagonal elements of $\op{\omega}_E$ in the basis $\mathcal B$.} (definition of von Neumann entropy as minimum of Shannon entropy ovel all measurements) and for PQC channels we have
${\rm diag}_{\mathcal B}[\op{\omega}_E]=\sum_k p_k {\rm Tr}[\op{U}_k \op{\rho}_A \du{\op{U}}_k]
\ket{k}\bra{k}$. Using the the trace properties and normalization of $\op{\rho}_A$ we obtain ${\rm diag}_{\mathcal B}[\op{\omega}_E]=\{p_k\}_k$, i.e. $S({\rm diag}_{\mathcal B}[\op{\omega}_E])=H(\{p_k\}_k)$.

In what follows we will show that for qubit the inequality is saturated for decomposition into orthogonal unitaries, i.e. for Pauli channels. From the previous paragraph it is clear that it is sufficient to show that for some $\op{\rho}$ the induced enviroment state $\op{\omega}_E$ is diagonal. Hence, we have to verify the conditions
under which the identity ${\rm Tr}[\op{U}_j\op{\rho} \du{\op{U}}_k]=0$ holds for $j\ne k$. In such case the
inequality is saturated. It is easy to see that by choosing $\op{\rho}=\frac{1}{2}\id$ this is the condition for orthogonality of transformations $\op{U}_j$ and this justifies our statement.

We have shown that for orthogonal decomposition of the channel $\sop{E}$ the entropy of the inequality is saturated,
i.e. entropy of the key equals to entropy exchange and this is indeed the maximal value of entropy exchange. Fortunately, the entropy exchange does not depend on the particular decomposition and therefore the entropy of the key cannot be lower for another decompositions. It turns out that for qubits any unital channel can be written as a convex
combination of orthogonal unitaries. However, for larger systems this is not the case in general. Consequently, the qubit PQC channel with minimal entropy of the key is the one with orthogonal encoding operations, i.e. the
corresponding Pauli channel $\Phi_\sop{E}$.

The necessary and sufficient entropy of the key is $1$ when $\op{\rho}^{(0)}=\overline{\rho}$ and it grows up to $2$ bits as the state $\op{\rho}^{(0)}$ approaches $\frac{1}{2}\id$. Therefore, it is natural to express the entropy as a function of the parameter
\begin{equation}
r=\frac{D(\op{\rho}^{(0)},1/2\id)}{D(\overline{\rho},1/2\id)},
\end{equation}
where the radius of the Bloch ball is $1$.

\begin{figure}
  \begin{center}
  \includegraphics[width=7cm]{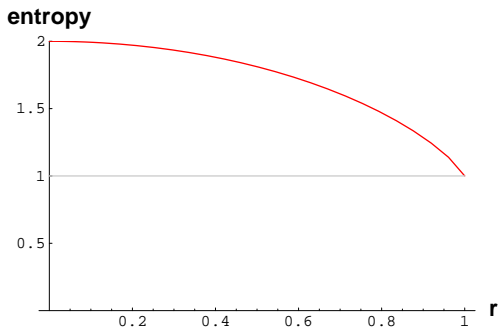}
  \caption{Dependency of the entropy on the variable $r$.}
  \label{PQC3d5}
  \end{center}
\end{figure}

Let us use the parametrization of
states $\op{\rho}_1,\op{\rho}_2,\op{\rho}_3\in\set{P}$ introduced in the Eq. \eqref{ksi}
and put $s=D(\op{\rho}^{(0)},1/2\id)$ and
$p=D(\overline{\rho},1/2\id)=|\alpha\beta|/\sqrt{1+\alpha^2+\beta^2}$.
Comparing it with the Eq. \eqref{equ:Pauli_chan_probab} we obtain that
the probabilites reads $p_0=p_z=\frac{1}{4}(1+r)$
 and $p_x=p_y=\frac{1}{4}(1-r)$, where we used the relation
 $r=s/p$. The evaluation of the entropy for this realization of PQC
 channel leads us to formula
\begin{equation}
\begin{split}
H(\{p_j\}_j)&=-\sum_j p_j\log_2 p_j=2-\frac{1}{2}\left[(1+r)\log_2(1+r)+(1-r)\log_2(1-r)\right],
\end{split}
\end{equation}
where $0\le r\le 1$. It is easy to see that $1\le H(\{p_j\}_j)\le
2$. The graph of the function $H(\{p_j\}_j)$ depending on the variable
$r$ is on the Figure \ref{PQC3d5}. Unfortunately, as we see from the
graph, the entropy grows very fast as $r$ goes to $0$. In example for
$r=1/2$ the entropy is already $H(\{p_j\}_j)\approx 1.81128$.

%%%%%%%%%%%%%%%%%%%%%%%%%%%%%%%%%%%%%%%%%%%%%%%%%%%%%%%%%%%%%%%%%%%%%%%%%%
\section{Conclusion}
\label{sec:Conclus}
%%%%%%%%%%%%%%%%%%%%%%%%%%%%%%%%%%%%%%%%%%%%%%%%%%%%%%%%%%%%%%%%%%%%%%%%%%

\subsection*{All single-qubit private quantum channels}
In this paragraph we will answer the following question:
which unital maps constitute a PQC? We have shown that it is
sufficient to consider only Pauli channels, i.e. the maps
$\Phi_\sop{E}={\rm diag}\{1,\lambda_1,\lambda_2,\lambda_3\}$.
Nontrivial private quantum channels are characterized by the property,
that at least two pure states $|\psi_1\rangle,|\psi_2\rangle$ are
mapped into the same state $\op{\rho}^{(0)}$. In the Bloch sphere
parametrization ($\vec{r}=(r_x,r_y,r_z)$)
this means that $\vec{r}_1\mapsto\vec{r}^\prime_1=\vec{s}$ and
$\vec{r}_2\mapsto\vec{r}_2^\prime=\vec{s}$.
Using these relations and explicit form of the
Pauli channel we come to the following ``PQC'' conditions
$0=\vec{r}_1^\prime-\vec{r}^\prime_2$, i.e.
$\lambda_j(r_{1j}-r_{2j})=0$ for all components $j=x,y,z$.
This equality is satisfied only if $\lambda_j=0$ for some $j$, or $r_{1j}=r_{2j}$.
Consider the case when none of the $\lambda$s vanishes,
i.e. $\lambda_1\lambda_2\lambda_3\ne 0$. It follows that in order to
fulfill the PQC conditions $r_{1j}=r_{2j}$ for every $j$. But it
means that the states are the same. Therefore, at least one of the
parameters $\lambda_j$ must vanish. Otherwise the transformation does
not correspond to private quantum channel. The complete positivity
condition restricts the possible values of $\lambda_1,\lambda_2$
(we put $\lambda_3=0$) so that the inequality
$|\lambda_1\pm\lambda_2|\le 1$
characterize all the possible qubit private quantum channels.

\subsection*{Multi-qubit generalization}
This result can be generalized for a specific class of multi-qubit
states. In each of the $n$ qubits we choose a set of plaintexts
$\overline{\set{P}}_k$ in the corresponding Bloch ball. Each of the
qubits can be encoded by PQC $\sop{E}_k:\overline{\set{P}}_k\mapsto\op{\rho}^{(0)}_k$.
Following the single qubit results, we design a PQC on each of the
qubits, which encrypts any state of the form
\begin{equation}
\sum_i \mu_i \rho_1^{(i)}\otimes\cdots\otimes\rho_n^{(i)},
\end{equation}
where $\forall i:\mu_i\in\mathbb R$, $\sum_i\mu_i=1$
and $\forall i,k:\rho_k^{(i)}\in \overline{\set{P}}_k$.
Let us denote the set of such states by $\overline{\set{P}}$. Note
note that this set contains entangled states as well, because
not only convex combinations of factorized states are allowed. The
values of $\mu_i$ are arbitrary. Consider for instance two qubits.
Using PQC encryption for $\set{P}={\mathcal S}(\H)$ on each qubit
enables us to encrypt each two-qubit (even entangled) quantum state.

\subsection*{Other implications}
In this paper we derived the restriction that the state $\op{\rho}^{(0)}$ of
the private quantum channel can be any state which has the distance
from the maximally mixed state $\frac{1}{2}\id$ the same or smaller than the
state $\overline{\rho}$, where $\overline{\rho}$
denotes the most mixed state in the linear span of $\set{P}$. We showed
that any of these states can be achieved
in the case of the qubit and therefore this condition is also sufficient.

Further, we demonstrated that it is enough to use a single bit of key
to encrypt the set $\overline{\set{P}}$, which is spanned by two linearly
independent states, and that any state of the previously described
candidates to the state $\op{\rho}^{(0)}$ can be achieved. We derived the same
result for the set $\overline{\set{P}}$ containing three linearly independent
states, but with the restriction that a single bit of the key suffices
provided that the state $\op{\rho}^{(0)}$ has the same distance from $1/2\id$ as
the state $\overline{\rho}$. As the distance of the state $\op{\rho}^{(0)}$ to $\frac{1}{2}\id$
approaches $0$, the necessary and sufficient entropy of the key approaches $2$.

As a special consequence of our derivation we obtain the result of \cite{Ambainis+Mosca...-Priva_quant_chann:2000} that the state $\op{\rho}^{(0)}=\frac{1}{2}\id$ when $\frac{1}{2}\id$ is in the convex span of $\set{P}$ and two bits of the key are needed to encrypt a qubit. Another special consequence of the above derivations is the result of \cite{Ambainis+Mosca...-Priva_quant_chann:2000} that to encrypt real combinations of two orthogonal basis states it is necessary and sufficient to use a single bit of key. These real combinations form a circle on the surface of the Bloch ball with center coinciding with the center of the Bloch ball.

Moreover, from the discussion in Section \ref{sub:Three-linearl-indepen-states} it follows that the impossibility of universal not operation on qubit \cite{Buzek+Hilery...-Optim_manip_qubit:1999,Buzek+Hilery...-Unive_gate:2000} can be derived from the fact that one bit of the key is not sufficient to encrypt a qubit.

\section*{Acknowledgements}
Support of the project GA\v CR GA201/01/0413 is acknowledged.
M.Z. acknowledges the support of the Slovak Academy of Sciences via the project CE-PI and of project INTAS (04-77-7289).

%%%%%%%%%%%%%%%%%%%%%%%%%%%%%%%%%%%%%%%%%%%%%%%%%%%%%%%%%%%%%%%%%%%%%%%%%%
%%%%%%%%%%%%%%%%%%%%%%%%%%%%%%%%%%%%%%%%%%%%%%%%%%%%%%%%%%%%%%%%%%%%%%%%%%
\appendix
%%%%%%%%%%%%%%%%%%%%%%%%%%%%%%%%%%%%%%%%%%%%%%%%%%%%%%%%%%%%%%%%%%%%%%%%%%
\section{Bloch sphere and qubit channels}
%%%%%%%%%%%%%%%%%%%%%%%%%%%%%%%%%%%%%%%%%%%%%%%%%%%%%%%%%%%%%%%%%%%%%%%%%%
Qubit (two-dimensional quantum system)
provides us a very simple and illustrative picture of
the state space.
Any state can be expressed as a linear combination
of the operators $\{\id,\op{\sigma}_x,\op{\sigma}_y,\op{\sigma}_z\}$.
In particular,
each operator $\op{\rho}=\frac{1}{2}(\id+\vec{r}\cdot\vec{\sigma})$
has a unit trace and if $|\vec{r}|\le 1$, then it is also positive.
Consequently, the state space forms a ball with the unit radius.
The equivalence $\op{\rho}\leftrightarrow\vec{r}$ is called {\it Bloch sphere representation}
(see for instance Refs.~\cite{Nielsen+Chuang-Quant_Compu_Quant:2000,Preskill-Lectu_notes_quant}).
From the orthogonality relation ${\rm Tr}\op{\sigma}_k\op{\sigma}_l=2\delta_{kl}$
the parameters of state are given by a simple formula
$\vec{r}={\rm Tr}\op{\rho}\vec{\sigma}$, i.e. as
the mean values of the hermitian operators (measurements)
$\op{\sigma}_x,\op{\sigma}_y,\op{\sigma}_z$.

Let us describe the relation between the density operators ${\cal S}({\cal H})$
(three-parametric subset) embedded in four-dimensional space of
Hermitian operators and the Bloch sphere contained in three-dimensional space.
Let us denote by $\op{\rho}_j$ ($j=1,2,3,4$) the basis of this space
corresponding to four density operators. The vectors $\vec{r}_j$
represents the associated points in the Bloch sphere (in three dimensional real vector space).
Only {\it trace-preserving linear combinations}, i.e.
$\op{\rho}=\sum_j a_j\op{\rho}_j$ with $\sum_j a_j=1$ for real $a_j$,
can be understand as linear combinations of the vectors within the
Bloch sphere picture, i.e. $\vec{r}=\sum_j a_j\vec{r}_j$. In fact,
Bloch sphere is situated in the three-dimensional space of Hermitian
operators with unit trace, but only special linear combinations
($\sum_j a_j=1$) of Bloch vectors has its counterparts in the original
space of Hermitian operators.

The structure of qubit channels is known mainly due to work
of Ruskai {\it et al.} \cite{King+Ruskai-Minim_entro_state:2001,Ruskai+Szarek...-analy_compl_trace:2001}.
Let us now briefly present a corresponding geometrical picture.
From the mathematical point
of view  \cite{Nielsen+Chuang-Quant_Compu_Quant:2000,Preskill-Lectu_notes_quant} the channels are described by
linear trace-preserving completely positive maps $\sop{E}$ defined on the
set of operators. The complete positivity is guaranteed
if the operator $\Omega_{\sop{E}}=(\sop{E}\otimes\id)\op{P}_+$ is a valid quantum state\footnote{$\op{P}_+$ is a projection onto maximally entangled state $\ket{\psi_+}=\frac{1}{\sqrt{2}}(\ket{00}+\ket{11})$.}.
Any qubit channel $\sop{E}$ can be illustrated as an affine transformation of
the Bloch vector $\vec{r}$, i.e. $\vec{r}\mapsto\vec{r}^\prime= \op{T}\vec{r}+\vec{t}$,
where $\op{T}$ is a real 3x3 matrix and $\vec{t}$ is a translation.
This form guarantees that the transformation $\sop{E}$ is hermitian and
trace preserving, but the complete positivity conditions define (nontrivial)
constraints on possible values of parameters. In fact, the set of all
completely positive trace-preserving maps forms a specific convex
subset of all affine transformations.

Any matrix $\op{T}$ can be written in the so-called
singular value decomposition, i.e.
$\op{T}=\op{R}_{\op{U}}\op{D}\op{R}_{\op{V}}$ with $\op{R}_{\op{U}},\op{R}_{\op{V}}$ corresponding to orthogonal rotations and $\op{D}={\rm diag}\{\lambda_1,\lambda_2,\lambda_3\}$ being diagonal
with $\lambda_k$ the singular values of $\op{T}$. This means that any
map $\sop{E}$ is a member of less-parametric family of maps
of the ``diagonal form'' $\Phi_{\sop{E}}$. In particular
$\sop{E}[\op{\rho}]=\op{U}\Phi_{\sop{E}}[\op{V}\op{\rho} \du{\op{V}}]\du{\op{U}}$ with
$\op{U},\op{V}$ unitary operators.
The reduction of parameters is very helpful,
and most of the properties (also complete positivity)
of $\sop{E}$ is reflected
by the properties of $\Phi_{\sop{E}}$. The map $\sop{E}$ is completely positive only if
$\Phi_{\sop{E}}$ is. Let us note that $\Phi_{\sop{E}}$ is determined not only by
the matrix $\op{D}$, but also by a new translation vector $\vec{\tau}=\op{R}_{\op{U}}\vec{t}$,
i.e. under the action of the map $\Phi_{\sop{E}}$ the
Bloch sphere transforms as follows $r_j\mapsto r_j^\prime=\lambda_j r_j+\tau_j$.

A special type of completely positive maps are the unital ones, i.e. those for which the total mixture (center of the Bloch sphere) is preserved. For these channels the translation term vanishes, $\vec{t}=\vec{\tau}=\vec{0}$, and the Bloch sphere is ``shrinked'' without shifting its center. In this case the analysis of all possible channels is quite simple, because the induced map $\Phi_{\sop{E}}$ is uniquely specified only by three real parameters. Positivity of the transformation $\Phi_{\sop{E}}$ corresponds to conditions $|\lambda_k|\le 1$, i.e. all points lying inside a cube. The conditions of complete positivity \cite{King+Ruskai-Minim_entro_state:2001,Ruskai+Szarek...-analy_compl_trace:2001} demands the validity of the following four inequalities $|\lambda_1\pm\lambda_2|\le|1\pm\lambda_3|$. This specifies the tetrahedron lying inside a cube of all positive unital maps with extremal points being four unitary transformations $\id,\op{\sigma}_x,\op{\sigma}_y,\op{\sigma}_z$.

It follows that each unital map is unitarily equivalent to
the map of the form $\Phi_{\sop{E}}={\rm diag}\{1,\lambda_1,\lambda_2,\lambda_3\}$.
The set of all unital channels is convex. Obviously the unitary channels
are extremal points of this set. Let us consider a Pauli channel
$\sop{P}[\op{\rho}]=\sum_k p_k \op{\sigma}_k\op{\rho}\op{\sigma}_k$, i.e. a general
convex combination of four Pauli unitary transformations. Rewriting this
action in the Bloch sphere parameters we obtain the transformation
\begin{equation}
\sop{P}=
\left(
\begin{array}{cccc}
 1 & 0 & 0 & 0 \\
 0 & 1-2(p_y+p_z) & 0 & 0 \\
 0 & 0 & 1-2(p_x+p_z) & 0 \\
 0 & 0 & 0 & 1-2(p_x+p_y)
\end{array}
\right).
\end{equation}
As a result we get that unital channels are unitarily equivalent to
Pauli channel. Consequently, each unital channel $\sop{E}$ can be written
as a convex combination of (at least) four unitary channels. The
probabilities are determined by the parameters of the induced map
$\Phi_{\sop{E}}$
\begin{equation}
\begin{array}{rcl}
p_x&=& \frac{1}{4}(1+\lambda_1-\lambda_2-\lambda_3) \\
p_y&=& \frac{1}{4}(1-\lambda_1+\lambda_2-\lambda_3) \\
p_z&=& \frac{1}{4}(1-\lambda_1-\lambda_2+\lambda_3) \\
p_0&=& 1-p_x-p_y-p_z.\\
\end{array}
\end{equation}

Unitary channels are rotations of the Bloch sphere. Unital channels
are rotations combined with the deformation so that the output states
form an ellipsoid centered in the total mixture. The values $\lambda_j$
define the size of the ellipsoid along three main axes.

%GATHER{../../../bibliografie/qcrypto.bib}
\bibliographystyle{plain}
\bibliography{../../../bibliografie/qcrypto}
\end{document}